\documentclass[aps,prd,showpacs,showkeys,amsmath,amssymb]{revtex4}
\usepackage{graphicx}
\usepackage{bm}
\begin{document}

\title{$D_{sJ}^+(2632)$: An Excellent Candidate of Tetraquarks}
\author{Y.-R. Liu}
\affiliation{Department of Physics, Peking University, Beijing
100871, China}
\author{Y.-B. Dai}
\author{C. Liu}
\affiliation{Institute of Theoretical Physics,
 Chinese Academy of Sciences, P.O. Box 2735, Beijing 100080,
 China}
\author{Shi-Lin Zhu}
\email{zhusl@th.phy.pku.edu.cn}
\affiliation{Department of
Physics, Peking University, Beijing 100871, China}

\date{\today}
\begin{abstract}

We analyze various possible interpretations of the narrow state
$D_{sJ}(2632)$ which lies 100 MeV above threshold. This
interesting state decays mainly into $D_s \eta$ instead of $D^0
K^+$. If this relative branching ratio is further confirmed by
other experimental groups, we point out that the identification of
$D_{sJ}(2632)$ either as a $c\bar s$ state or more generally as a
${\bf {\bar 3}}$ state in the $SU(3)_F$ representation is probably
problematic. Instead, such an anomalous decay pattern strongly
indicates $D_{sJ}(2632)$ is a four quark state in the $SU(3)_F$
${\bf 15}$ representation with the quark content ${1\over
2\sqrt{2}}
(ds\bar{d}+sd\bar{d}+su\bar{u}+us\bar{u}-2ss\bar{s})\bar{c}$. We
discuss its partners in the same multiplet, and the similar
four-quark states composed of a bottom quark $B_{sJ}^0(5832)$.
Experimental searches of other members especially those exotic
ones are strongly called for.

\end{abstract}

\maketitle

\pagenumbering{arabic}

\section{Introduction}

The experimental discovery of the low-lying narrow charm mesons
$D_{sJ}(2317)$ and $D_{sJ}(2457)$ \cite{babar,cleo,belle1,belle2}
has attracted much attention. Since their masses in the
constituent quark model are roughly 160 MeV higher than the
experimental values, some people postulated these states could be
a four quark state \cite{cheng,rev1}. However, there is no
compelling evidence that $D_{sj}(2317)$ and $D_{sJ}(2457)$ are
non-conventional meson states. It is still possible to interprete
them as $c\bar s$ states \cite{bardeen,dai}. Instead, the large
electromagnetic branching ratio of $D_{sJ}(2457)$ favors such a
picture \cite{belle1,belle2}. Interested readers may consult Ref.
\cite{rev2} for a nice review of this topic.

Very recently SELEX Collaboration observed another exotic
charm-strange meson $D_{sJ}(2632)$ with a significance of
$7.2\sigma$ in the $D_s \eta$ channel and $5.3\sigma$ in the
$D^0K^+$ channel \cite{selex}, which has created some excitement
\cite{mai,chen,ni,zhao,barnes}. The decay width of this narrow
resonance is less than 17 MeV at $90\%$ confidence level.

At present there are three puzzles concerning this state. First,
this state lies 274 MeV above $D^0K^+$ threshold and 116 MeV above
$D_s \eta$ threshold. One would naively expect its strong decay
width to be around $(100\sim 200)$ MeV. Both particles in the
final states are pseudoscalar mesons, thus $D_{sJ}^+(2632)$ has
$J=L, P=(-)^L$ with $L$ being the decay angular momentum. Because
the final states $D_s^+$ and $\eta$ are both isoscalar,
$D_{sJ}^+(2632)$ is probably an isoscalar.

Secondly, the ground state charm-strange mesons with $L=0$ are
$D_s(1968)$ and $D_s^\ast (2112)$. According to the heavy quark
effective field theory, there exist two heavy doublets with
positive parity and $L=1$. We denote them as $l^P={1\over 2}^+$
and $l^P={3\over 2}^+$ where $l$ is the angular momentum of the
light quark. The $l^P={1\over 2}^+$ doublet are $D_{sJ}(2317)$ and
$D_{sJ}(2457)$. The $l^P={3\over 2}^+$ doublet are $D_{sJ}(2536)$
and $D_{sJ}(2573)$ \cite{pdg}. One may be tempted to interpret
$D_{sJ}(2632)$ either as a member of the $l^P={5\over 2}^-$
doublet with $J=3, L=2$ or as a member of the $l^P={3\over 2}^-$
doublet with $J=1, L=2$. Especially the identification of
$D_{sJ}(2632)$ as a $J=3, L=2$ state may seem attractive at first
sight since the presence of the high angular momentum may lead to
a small decay width. But apparently $D_{sJ}(2632)$ is too low for
$L=2$! Another possibility is that $D_{sJ}(2632)$ is the first
radial excitation of the ground state charm-strange meson
\cite{zhao}.

The most demanding issue is the unusual decay pattern. This state
decays mainly into the $D_s \eta$ mode. Recall that SU(3) flavor
symmetry breaking is at most around $20\%$ and the physical eta
meson is mainly an octet. One may perform a more refined analysis
taking into account the mixing between $\eta_8$ and $\eta_1$ or
the SU(3) symmetry breaking effects. But the following result will
not change dramatically. We may write an effective Lagrangian for
$D_{sJ}(2632)$ decay processes if it is a $c\bar s$ state. The
decay width reads
\begin{equation}\label{ratio}
\Gamma=\lambda^2 g^2\frac{k^{2L+1}}{m^{2L}},
\end{equation}
where $\lambda$ is the C.G. coefficient, $g$ is the universal and
dimension-less effective coupling constant, $L$ is the angular
momentum for decay. $m$ is the parent mass and $k$ is the decay
momentum in the center of mass frame
\begin{equation}
k=\frac{1}{2m} \{[m^2-(m_1+m_2)^2][m^2-(m_1-m_2)^2]\}^{\frac12},
\end{equation}
where $m_1$ and $m_2$ are the masses of final mesons. The ratio of
decay widths of these two channels is
\begin{equation}
\frac{\Gamma (D^0 K^+)}{\Gamma (D_s\eta)}=(\frac{\lambda_{D^0
K^+}}{\lambda_{D_s\eta}})^2(\frac{k_{D^0
K^+}}{k_{D_s\eta}})^{2L+1}.
\end{equation}
Using $\frac{\lambda_{D^0 K^+}}{\lambda_{D_s\eta}}=\sqrt{3\over
2}$, $k_{D^0 K^+}=499$ MeV, $k_{D_s\eta}=325$ MeV, we get
\begin{equation}
\frac{\Gamma (D^0 K^+)}{\Gamma (D_s\eta)}=2.3*(1.54)^{2L}\ge 2.3.
\end{equation}
which is nearly 15 times larger than the experimental value
\cite{selex}
\begin{equation} \frac{\Gamma (D^0 K^+)}{\Gamma
(D_s\eta)}=0.16\pm 0.06 \; .
\end{equation}
If this relative branching ratio is confirmed by other
experimental groups, we conclude that the identification of
$D_{sJ}(2632)$ either as a $c\bar s$ state or more generally as a
${\bf {\bar 3}}$ state in the $SU(3)_F$ representation is very
problematic. We must seek other interpretations.

$D_{sJ}(2632)$ decays mainly into $D_s \eta$. $\eta$ is a mixture
of $\eta_8$ and $\eta_1$. $\eta_1$ is a SU(3) singlet which mixes
strongly with $G\tilde G$. One may think $D_{sJ}(2632)$ is a good
candidate of heavy hybrid meson with the content $c G \bar s$
\cite{zhu}. If it is the lowest hybrid state, one should expect
that it is composed of $c, \bar s$ and a gluon of magnetic field
type (all in S state) so that it has spin-parity $1^-$. For decay
of this state to $D_s$ and $\eta$, the quark should emit a gluon
of electric field type in S state so that the gluon component has
total spin-parity $0^-$. However, in this transition the quark
component must jump to the excited state due to selection rule
contrary to the experiment observation. If the initial state is
not the lowest hybrid state, its mass should be heavier than the
observed mass. The hybrid state with two explicit gluons $G\tilde
G$ also seems to be too heavy. However, the gluon is a flavor
singlet. So the hybrid assumption can not explain the unusual
decay pattern.

Since $D_{sJ}(2632)$ is above threshold, it can't be a hadron
molecule state. Molecules are bound states of color singlet
hadrons. They should lie near or below threshold. However, it
could be a bound state of two color non-singlet clusters like a
diquark and anti-diquark. That's what we will advocate below:
$D_{sJ}(2632)$ is a four quark state which provides a simple and
natural explanation of the unusual decay pattern.

%%%%%%%%%%%%%%%%%%%%%%%%%%%%%%%%%%%%%%%%%%%%%%%%%%%
\section{Tetraquark multiplets}
%%%%%%%%%%%%%%%%%%%%%%%%%%%%%%%%%%%%%%%%%%%%%%%%%%%

In this section, we consider a tetraquark state $qq\bar{q}\bar{c}$
with $q=u,d, s$. We present the wave functions and the decay modes
of the tetraquark states with one anti-charm quark.

Under the transformation of $SU(3)_F$, the charm quark is singlet.
There are four multiplets according to
\begin{equation}
3\otimes3\otimes\bar{3}\otimes1=3\oplus3\oplus\bar{6}\oplus15 \; .
\end{equation}

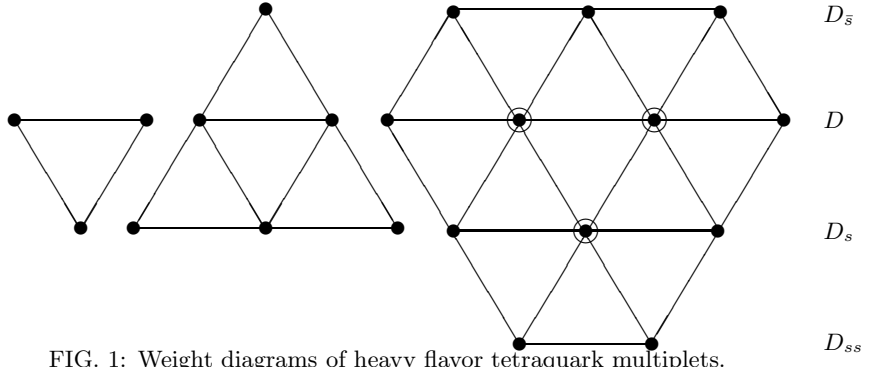
\begin{figure}[h]
\begin{center}
\begin{picture}(150,100)

\thinlines
%   triplet
\put(-66,63){\line(1,0){50}}\put(-66,63){\line(3,-5){25}}
\put(-16,63){\line(-3,-5){25}}

\put(-66,63){\circle*{5}}\put(-16,63){\circle*{5}}
\put(-41,22){\circle*{5}}
%   anti-sextet
\put(4,63){\line(1,0){50}}
\put(29,105){\line(-3,-5){50}}\put(29,105){\line(3,-5){50}}
\put(-21,22){\line(1,0){100}} \put(4,63){\line(3,-5){25}}
\put(54,63){\line(-3,-5){25}}

\put(29,105){\circle*{5}} \put(4,63){\circle*{5}}
\put(54,63){\circle*{5}} \put(-21,22){\circle*{5}}
\put(29,22){\circle*{5}} \put(79,22){\circle*{5}}
%\put(95,22){\line(3,5){50}}

% 15-plet
\put(74,62){\line(3,5){25}} \put(100,105){\line(1,0){100}}
\put(200,105){\line(3,-5){25}}
\put(225,62){\line(-3,-5){50}}\put(175,-22){\line(-1,0){50}}\put(125,-22){\line(-3,5){50}}
\put(74,63){\line(1,0){150}} \put(100,20){\line(3,5){50}}
\put(100,21){\line(1,0){100}} \put(200,22){\line(-3,5){50}}
\put(125,-22){\line(3,5){75}}\put(175,-22){\line(-3,5){75}}

\put(100,104){\circle*{5}}\put(151,104){\circle*{5}}\put(201,104){\circle*{5}}
\put(75,63){\circle*{5}}\put(125,63){\circle*{5}}
\put(176,63){\circle*{5}}\put(225,63){\circle*{5}}
\put(125,63){\circle{9}}\put(176,63){\circle{9}}
\put(100,21){\circle*{5}}\put(150,21){\circle*{5}}\put(200,21){\circle*{5}}
\put(150,21){\circle{9}}
\put(125,-22){\circle*{5}}\put(175,-22){\circle*{5}}

\put(240,100){$D_{\bar{s}}$} \put(240,60){$D$} \put(240,18){$D_s$}
\put(240,-24){$D_{ss}$}
\end{picture}
\end{center}
\caption{Weight diagrams of heavy flavor tetraquark
multiplets.}\label{fig2}
\end{figure}

In these tetraquark multiplets, all states have charm number
$C=-1$. We denote the states with strangeness $S=1$ as
$D_{\bar{s}}$, with $S=0$ as $D$, with $S=-1$ as $D_s$ and with
$S=-2$ as $D_{ss}$. The weight diagrams are shown in Fig.
\ref{fig2}.

We present all the tetraquark flavor wave functions in Table
\ref{tab3}-\ref{tab5}. These expressions are obtained by operating
$U_-$ and $I_-$ operators on the highest weight state.

\begin{table}[h]
\begin{center}
\begin{tabular}{c|c|c||c|c}\hline
&{States} &Flavor Wave Functions&{States} &Flavor Wave Functions\\
\hline
$I=\frac12,S=0$ & $\bar{D}_3^0$&$\frac12(su\bar{s}-us\bar{s}-ud\bar{d}+du\bar{d})\bar{c}$&$\bar{D}_3^{0\prime}$&$\frac{1}{2\sqrt2}(su\bar{s}+us\bar{s}+ud\bar{d}+du\bar{d}+2uu\bar{u})\bar{c}$ \\
  & $D_3^-$&$\frac12(ud\bar{u}-du\bar{u}-ds\bar{s}+sd\bar{s})\bar{c}$&$D_3^{-\prime}$&$\frac{1}{2\sqrt2}(ud\bar{u}+du\bar{u}+ds\bar{s}+sd\bar{s}+2dd\bar{d})\bar{c}$ \\
  \hline
$I=0,S=-1$ & $D_{s,3}^-$&$\frac{1}{2}(ds\bar{d}-sd\bar{d}-su\bar{u}+us\bar{u})\bar{c}$&$D_{s,3}^{-\prime} $&$\frac{1}{2\sqrt2}(ds\bar{d}+sd\bar{d}+su\bar{u}+us\bar{u}+2ss\bar{s})\bar{c}$\\
\hline
\end{tabular}
\end{center}
\caption{The flavor wave functions of the heavy tetraquarks in the
two triplets.} \label{tab3}
\end{table}

\begin{table}[h]
\begin{center}
\begin{tabular}{c|c|c}\hline
&{States} &Flavor Wave Functions\\
\hline
$I=0,S=1$ & $D_{\bar{s},\bar{6}}^0$ &  $\frac{1}{\sqrt2}(ud\bar{s}-du\bar{s})\bar{c}$        \\
\hline
$I=\frac12,S=0$ & $D_{\bar{6}}^0$ & $\frac12(us\bar{s}-su\bar{s}-ud\bar{d}+du\bar{d})\bar{c}$ \\
  & $D_{\bar{6}}^-$ & $\frac{1}{2}(ds\bar{s}-sd\bar{s}+ud\bar{u}-du\bar{u})\bar{c}$ \\
  \hline
$I=1,S=-1$ & $D_{s,\bar{6}}^0$ & $\frac{1}{\sqrt2}(su\bar{d}-us\bar{d})\bar{c}$\\
 & $D_{s,\bar{6}}^-$ & $\frac12(sd\bar{d}-ds\bar{d}-su\bar{u}+us\bar{u})\bar{c}$\\
  &$D_{s,\bar{6}}^{--}$ &  $\frac{1}{\sqrt2}(ds\bar{u}-sd\bar{u})\bar{c}$\\
\hline
\end{tabular}
\end{center}
\caption{The flavor wave functions of the heavy tetraquarks in the
anti-sextet $T_{ij}$.} \label{tab4}
\end{table}

\begin{table}[h]
\begin{center}
\begin{tabular}{c|c|c}\hline
&{States} &Flavor Wave Functions\\
\hline
$I=1,S=1$&$D^+_{\bar{s},15}$& $uu\bar{s}\bar{c}$\\
&$D^0_{\bar{s},15}$&$\frac{1}{\sqrt2}(ud\bar{s}+du\bar{s})\bar{c}$\\
&$D^-_{\bar{s},15}$&$dd\bar{s}\bar{c}$\\
 \hline
$I=\frac32,S=0$&$D^+_{15}$&$ -uu\bar{d}\bar{c}$\\
&${D}^0_{15}$&$ -\frac{1}{\sqrt3}(ud\bar{d}+du\bar{d}-uu\bar{u})\bar{c}$\\
&$D^-_{15}$&$\frac{1}{\sqrt3}(ud\bar{u}+du\bar{u}-dd\bar{d})\bar{c}$\\
&$D^{--}_{15}$&$ dd\bar{u}\bar{c}$\\
 \hline
$I=\frac12,S=0$&$D^{0\prime}_{15}$&$ \frac{1}{2\sqrt6}[3us\bar{s}+3su\bar{s}-ud\bar{d}-du\bar{d}-2uu\bar{u}]\bar{c}$\\
&$D^{-\prime}_{15}$&$\frac{1}{2\sqrt6}[3ds\bar{s}+3sd\bar{s}-ud\bar{u}-du\bar{u}-2dd\bar{d}]\bar{c}$\\
 \hline
$I=1,S=-1$&$D^0_{s,15}$&$-\frac{1}{\sqrt2}(su\bar{d}+us\bar{d})\bar{c}$\\
&$D^-_{s,15}$&$-\frac{1}{2}(ds\bar{d}+sd\bar{d}-su\bar{u}-us\bar{u})\bar{c}$\\
&$D^{--}_{s,15}$&$\frac{1}{\sqrt2}(ds\bar{u}+sd\bar{u})\bar{c}$\\
 \hline
$I=0,S=-1$&$D^{-\prime}_{s,15}$&$-\frac{1}{2\sqrt2}(ds\bar{d}+sd\bar{d}+su\bar{u}+us\bar{u}-2ss\bar{s})\bar{c}$\\
 \hline
$I=\frac12,S=0$&$D^-_{ss}$&$ -ss\bar{d}\bar{c}$\\
&$D^{--}_{ss}$& $ss\bar{u}\bar{c}$\\
 \hline
\end{tabular}
\end{center}
\caption{The flavor wave functions of the heavy tetraquarks in
$15$-plet $T^{ij}_k$.} \label{tab5}
\end{table}

As we have pointed out, $D_{sJ}(2632)$ can not be a member of
$\bar 3$ representation. From now on, we focus on the ${\bar 6},
15$ representations. We can get the decay modes of these
tetraquark states by expanding the following effective Lagrangian
\begin{eqnarray}\label{ham}
L_{eff}=g_6 T^{\dagger ij} M^a_i T^b \epsilon_{abj}+ g_{15}
T^{\dagger i}_{jm} M^j_i T^m \; ,
\end{eqnarray}
where the triplet $T^i$ reads
\begin{equation}\label{triplet}
(T^i)=\bordermatrix{&\cr & \bar{c}u\cr& \bar{c}d\cr
&\bar{c}s}=\bordermatrix{&\cr & \bar{D}^0\cr& D^-\cr &D^-_s} \; .
\end{equation}

The tensor representation for anti-sextet is
\begin{eqnarray}
(T_{ij})&=&\left(\begin{array}{ccc}
D^{--}_{s,\bar{6}}&-\frac{1}{\sqrt{2}}{D^-_{s,\bar{6}}}&\frac{1}{\sqrt{2}}{D^-_{\bar{6}}}\\
-\frac{1}{\sqrt{2}}{D^-_{s,\bar{6}}}&D^0_{s,\bar{6}}&-\frac{1}{\sqrt{2}}{{D}^0_{\bar{6}}}\\
\frac{1}{\sqrt{2}}{D^-_{\bar{6}}}&-\frac{1}{\sqrt{2}}{{D}^0_{\bar{6}}}&D^0_{\bar{s},\bar{6}}\end{array}\right).
\end{eqnarray}

The members of 15 representation are\\
\textbullet\ $Y = \frac43, I=1,$
\begin{eqnarray}
 T^{11}_{3}=D_{\bar{s},15}^+ \qquad
T^{12}_{3}=\frac{1}{\sqrt2}D_{\bar{s},15}^0 \qquad
T^{22}_{3}=D_{\bar{s},15}^-
\end{eqnarray}
\textbullet\ $Y = \frac13, I=3/2,1/2$
\begin{eqnarray}
 T^{11}_{2}=-D_{15}^+ \qquad
T^{11}_{1}=\frac{1}{\sqrt3}D_{15}^0-\frac{1}{\sqrt6}D_{15}^{0\prime}
\qquad
T^{12}_{2}=-\frac{1}{\sqrt3}D_{15}^0-\frac{1}{2\sqrt6}D_{15}^{0\prime}\qquad
T^{13}_{3}=\frac{\sqrt6}{4}D_{15}^{0\prime} \\
T^{12}_{1}=\frac{1}{\sqrt3}D_{15}^--\frac{1}{2\sqrt6}D_{15}^{-\prime}\qquad
T^{22}_{2}=-\frac{1}{\sqrt3}D_{15}^-
-\frac{1}{\sqrt6}D_{15}^{-\prime} \qquad
T^{23}_{3}=\frac{\sqrt6}{4}D_{15}^{-\prime} \qquad
T^{22}_{1}=D_{15}^{--}
\end{eqnarray}
\textbullet\ $Y = -\frac23, I=1,0$
\begin{eqnarray}
&T^{13}_{2}=-\frac{1}{\sqrt2}D_{s,15}^0 \qquad T^{13}_{1}=\frac12
D_{s,15}^--\frac{\sqrt2}{4} D_{s,15}^{-\prime}\qquad
T^{23}_2=-\frac12 D_{s,15}^--\frac{\sqrt2}{4}
D_{s,15}^{-\prime}&\nonumber\\
&T^{33}_{3}=\frac{1}{\sqrt2}D_{s,15}^{-\prime} \qquad
T^{23}_{1}=\frac{1}{\sqrt2}D_{s,15}^{--}&
\end{eqnarray}
\textbullet\ $Y = -\frac53, I=1/2$
\begin{eqnarray}
T^{33}_{2}=-D_{ss}^-\qquad T^{33}_{1}=D_{ss}^{--}.
\end{eqnarray}
These expressions can also be obtained by using the isospin and U
spin lowering operators.

We present the C.G. coefficient of each interaction term in Table
\ref{tab1} and \ref{tab2}. With these C.G. coefficients, it is
easy to derive the relative branching ratio.

\begin{table}[h]
\begin{center}
\begin{tabular}{cc|cc|cc}\hline
\multicolumn{2}{c}{$D_{s,\bar{6}}^{--}$} &
\multicolumn{2}{c}{$D_{s,\bar{6}}^-$} &
\multicolumn{2}{c}{$D_{s,\bar{6}}^0$}
\\ \hline
$K^- D^{-}$ & $-1$ & $K^- \bar{D}^{0}$ & $-\frac{1}{\sqrt2}$
&$\pi^+
D_s^{-}$ & $-1$\\
 $\pi^- D_s^{-}$ & $1$ &
$\bar{K^0}D^{-}$ & $\frac{1}{\sqrt{2}}$ &
$\bar{K^0}\bar{D}^{0}$&$1$\\
 & & $\pi^0 D_s^{-}$ & $1$ & & \\
\hline \multicolumn{2}{c}{$D_{\bar{6}}^-$} &
\multicolumn{2}{c}{$D_{\bar{6}}^0$}&
\multicolumn{2}{c}{$D_{\bar{s},\bar{6}}^0$}
\\ \hline
$\pi^-\bar{D}^{0}$ & $-\frac{1}{\sqrt2} $ & $\pi^0\bar{D}^{0}$ &
$-\frac{1}{2}$ & $K^0\bar{D}^{0}$ & $-1$  \\
$\pi^0 D^{-}$ & $\frac{1}{2}$ & $\eta_8\bar{D}^{0}$ &
$\frac{\sqrt3}{2}$
& $K^+ D^{-}$ & $1$ \\
$\eta_8D^{-}$ & $\frac{\sqrt3}{2}$ & $\pi^+D^{-}$ &
$-\frac{1}{\sqrt2}$ & &  \\
$K^0 D_s^{-}$&$\frac{1}{\sqrt2}$ & $K^+ D_s^{-}$ &
$\frac{1}{\sqrt2}$ && \\
\hline
\end{tabular}
\end{center}
\caption{Couplings of the heavy tetraquark anti-sextet $T_{ij}$
with the heavy meson triplet $T^i$ and pseudoscalar meson octet
$M^i_j$. The universal coupling constant is omitted.} \label{tab1}
\end{table}

\begin{table}[h]
\begin{center}
\begin{tabular}{cc|cc|cc|cc|cc}\hline
\multicolumn{2}{c}{$D^{+}_{15}$} &
\multicolumn{2}{c}{$D^{0}_{15}$} &
\multicolumn{2}{c}{$D^{-}_{15}$} &
\multicolumn{2}{c}{$D^{--}_{15}$} &
\multicolumn{2}{c}{$D^{0\prime}_{15}$}
\\ \hline
$\pi^{ +} \bar D^0$ & $-1$ & $\pi^{0} \bar D^0$ &
$\frac{2}{\sqrt{6}}$ & $\pi^{0}  D^-$ & $\frac{2}{\sqrt6}$ &
$\pi^{ -}  D^{-}$ & $1$ & $K^+
 D^-_s$ &$\frac{3}{2\sqrt6}$\\
& &$\pi^{ +}  D^-$ & $-\frac{1}{\sqrt3}$ & $\pi^{-} \bar D^0$
&$\frac{1}{\sqrt3}$& &&$\pi^+D^-$
&$-\frac{1}{2\sqrt6}$ \\
&&&&&&&& $\pi^{0}\bar D^0$ &$-\frac{1}{4\sqrt3}$\\
&&&&&&&& $\eta_8 \bar D^0 $ &
$-\frac34$  \\

\hline \multicolumn{2}{c}{$D_{15}^{-\prime}$} &
\multicolumn{2}{c}{$D_{s,15}^0$} &
\multicolumn{2}{c}{$D_{s,15}^-$} &
\multicolumn{2}{c}{$D_{s,15}^{--}$}&\multicolumn{2}{c}{$D_{s,15}^{-\prime}$} \\
\hline $K^0 D_s^-$ & $\frac{3}{2\sqrt6}$& $\pi^{+}D^-_s$ &
$-\frac{1}{\sqrt2}$ & $\pi^{0}D^-_s$ & $\frac{1}{\sqrt2}$ &
$\pi^{-}D^-_s$ & $\frac{1}{\sqrt2}$&$\eta_8 D^-_s$ &
$-\frac{\sqrt3}{2}$\\
 $\eta_8 D^-$ & $-\frac34$& $\bar{K}^{0} \bar D^0$
& $-\frac{1}{\sqrt2}$ & $K^- \bar D^0$ & $\frac12$ & $K^-
D^-$ & $\frac{1}{\sqrt2}$&$K^- \bar D^0$ & $-\frac{1}{2\sqrt2}$ \\
$\pi^{0} D^-$ & $\frac{1}{4\sqrt3}$& &  &$\bar{K}^0 D^-$ &
$-\frac12$   & &&$\bar{K}^{0}  D^-$ & $-\frac{1}{2\sqrt2}$ \\
$\pi^{-} \bar D^0$ & $-\frac{1}{2\sqrt6}$& &&&&&\\
 \hline
\multicolumn{2}{c}{$D_{\bar{s},15}^+$} &
\multicolumn{2}{c}{$D_{\bar{s},15}^0$} &
\multicolumn{2}{c}{$D_{\bar{s},15}^-$} &
\multicolumn{2}{c}{$D_{ss}^-$} &
\multicolumn{2}{c}{$D_{ss}^{--}$}\\
\hline $K^+ \bar D^0$ &$1$&  $K^+ D^-$ &$\frac{1}{\sqrt2}$&$K^0
D^-$ &$1$ &
$\bar{K}^0 D^-_s$ &$-1$ &$K^{-}D^-_s$ &$1$  \\
&&$K^0 \bar D^0$ &$\frac{1}{\sqrt2}$&&&& \\
\hline
\end{tabular}
\end{center}
\caption{Couplings of the heavy tetraquark $15$-plet $T^{ij}_k$
with usual meson octet $M^i_j$ and the heavy meson triplet $T^i$.
The universal coupling constant is omitted.} \label{tab2}
\end{table}

It is important to note that there is no isoscalar state with
valence quark content $\bar c s$ in the anti-sextet. However, in
the 15 tetraquark multiplet, there are an isoscalar
$D_{s,15}^{-\prime}$. From Table \ref{tab2}, we find the ratio of
C.G. coefficients for $\bar{D}^0 K^-$ channel and $D_s^- \eta_8$
channel is $\frac{1}{\sqrt6}$. Now the relative branching ratio
reads
\begin{equation}
{\Gamma(D_{s,15}^{-\prime}\rightarrow\bar{D}^0 K^-)\over
{\Gamma(D_{s,15}^{-\prime}\rightarrow D_s^-\eta)}}
=0.25*(1.54)^{2L}
\end{equation}
if we roughly assume $\eta\approx \eta_8$. This ratio is around
0.25 if $D_{s,15}^{-\prime}$ decays via S wave, which is
consistent with the experimental result $0.16\pm 0.06$. With P
wave decay the relative ratio is about 0.59.

The above derivation of the ratio of the SU(3) Clebsch-Gordan
coefficients is rather tedious. A straightforward way is to find
these coefficients from available tables in literature. An
exhaustive compilation of them is presented in Ref. \cite{cg}.
When a particle A in the SU(3) representation $R$ with hypercharge
$Y$, isospin $I, I_3$ couples to particle B $r, y, i, i_3$ and
particle C $r^\prime, y^\prime, i^\prime, i^\prime_3$, the SU(3)
Clebsch-Gordan coefficient can be decomposed into the product of
an iso-scalar part and SU(2) Clebsch-Gordan coefficient \cite{cg}:
 \begin{equation}
    \left< {\bf R} \, Y \, I \, I_3 | {\bf r} \, y \, i \, i_3 \,
                                      {\bf r'} \, y' \, i' \, i'_3 \right>
        = F({\bf R}, Y, I; {\bf r}, y, i, {\bf r'}, y', i')
        \times \left< I \, I_3 | i \, i_3 \, i' \, i'_3 \right> \;
        .
  \end{equation}
$F({\bf R}, Y, I; {\bf r}, y, i, {\bf r'}, y', i')$ is the
iso-scalar function which cab be found in the tables of Ref.
\cite{cg}. For example, from Table 10 in Ref. \cite{cg} we get
\begin{equation}
F(D_{s,15}^{-\prime} \to D_s \eta) ={\sqrt{3}\over 2}\; ,
\end{equation}
\begin{equation}
F(D_{s,15}^{-\prime} \to K^- {\bar D}^0) =-{1\over 2}\; .
\end{equation}
The SU(2) Clebsch-Gordan coefficient is $1$ for
$D_{s,15}^{-\prime} \to D_s \eta$ and ${1\over \sqrt{2}}$ for
$D_{s,15}^{-\prime} \to K^- {\bar D}^0$. Putting everything
together we obtain the relative ratio of the decay C. G.
coefficients again
\begin{equation}
|{\lambda (D_{s,15}^{-\prime} \to K^- {\bar D}^0)\over \lambda
(D_{s,15}^{-\prime} \to D_s \eta)}| ={1\over \sqrt{6}}\; .
\end{equation}

In fact, an even more transparent derivation of the relative ratio
of the decay coupling constants is possible if we assume the
"fall-apart" decay mechanism for tetraquarks. With the flavor wave
function of $D_{s,15}^{-\prime}$ in Table III, only the first two
terms contribute to $K^- {\bar D}^0$ decay mode while every piece
contributes to $D_s \eta $ mode. The ratio of their coupling
constants is simply
\begin{equation}
|\lambda (D_{s,15}^{-\prime} \to K^- {\bar D}^0):  \lambda
(D_{s,15}^{-\prime} \to D_s \eta)| =(1+1):{1\over
\sqrt{6}}\left(1+1+1+1+(-2)*(-2)*2\right)=1:\sqrt{6}\; .
\end{equation}
The last factor two arises because there are two possible ways to
get an $s\bar s$ from $ss\bar s$. Once again we reproduce this
ratio.

Based on the above argument, we propose that $D_{sJ}^+(2632)$ is
very probably a tetraquark state in the $SU(3)_F$ 15
representation with the quantum number $I=0, J^P=0^+$.

%%%%%%%%%%%%%%%%%%%%%%%%%%%%%%%%%%%%%%%%%%%%%%%%%%%%%
\section{Discussions}
%%%%%%%%%%%%%%%%%%%%%%%%%%%%%%%%%%%%%%%%%%%%%%%%%%%%%

We may also perform a rough estimate of the mass of
$D_{s,15}^{-\prime}$ with the flavor wave function. If we use
constituent quark masses: $m_u=m_d=310$ MeV, $m_s=450 MeV$ and
$m_c=1430$ MeV, we have
\begin{equation}
M_{D_{s,15}^{-\prime}}=\frac18[4(m_u+m_u+m_s)+4(m_s+m_s+m_s)]+m_c=m_u+2m_s+m_c=2640
\textrm{MeV}.
\end{equation}

Using the well-known Gell-Mann-Okubo formula, it's easy to derive
the mass relations between the tetraquark states within the same
multiplet
\begin{equation}
M=a+b[I(I+1)-\frac14 Y^2]+cY .
\end{equation}

For the heavy tetraquark anti-sextet,
\begin{equation}
M_{D_{\bar{s},\bar{6}}}-M_{D_{\bar{6}}}=M_{D_{\bar{6}}}-M_{D_{ss}}.
\end{equation}

For the heavy tetraquark 15-plet, we find
\begin{subequations}
\begin{eqnarray}
M_{D_{\bar{s},15}}-M_{D^\prime_{15}}& = & M_{D^\prime_{15}}-M_{D^\prime_{s,15}}\\
M_{D_{15}}-M_{D_{s,15}}& = & M_{D_{s,15}}-M_{D_{ss}}\\
M_{D_{\bar{s},15}}+3M_{D_{s,15}}& =
&2(M_{D_{15}}+M_{D^\prime_{s,15}}).
\end{eqnarray}
\end{subequations}

It is very interesting to note that there are three manifestly
exotic four quark states: $D^0_{\bar s, \bar 6}, D^0_{s, \bar 6},
D^{--}_{s, \bar 6}$. There are nine manifestly exotic four-quark
states: $D^+_{\bar s, 15}, D^0_{\bar s, 15}, D^-_{\bar s, 15},
D^+_{15}, D^{--}_{15}, D^0_{s, 15}, D^{--}_{s, 15}, D^-_{ss},
D^{--}_{ss}$. If $D_{sJ}(2632)$ is really a member of the
tetraquark states in the 15 representation, these exotic states
should appear as partners of $D_{sJ}(2632)$. We strongly call for
experimental search of these interesting states. Some of them have
a unique decay channel such as $D^+_{15}\to \bar{D}^0 \pi^+$,
$D^{--}_{15}\to D^- \pi^-$ etc.

Replacing the charm quark in $D^+_{sJ}(2632)$ by the bottom quark,
we get bottom tetraquark state $B^0_{sJ}(5832)$ where we have
assumed the mass difference between $D^+_{sJ}(2632)$ and
$B^0_{sJ}(5832)$ is simply $m_b-m_c=3200$ MeV. $B^0_{sJ}(5832)$ is
also expected to be a narrow resonance above threshold with a
dominant decay mode $B_s \eta$.

In Ref. \cite{mai} Maiani et al. suggested that $D^+_{sJ}(2632)$
is a $cd\bar d \bar s$ state to a good extent while its anomalous
decay ratio is caused by isospin breaking. In Ref. \cite{ni} it
was also suggested that $D^+_{sJ}(2632)$ is a cryptoexotic
tetraquark baryonium state $cd\bar d \bar s$ \cite{ni}. Ref.
\cite{chen} tried to identify all three narrow charm-strange
mesons $D_{sJ}(2137)$, $D_{sJ}(2457)$, $D^+_{sJ}(2632)$ as
four-quark states. Ref. \cite{zhao} suggested that
$D^+_{sJ}(2632)$ could be the first radial excitation of $D_s^\ast
(2112)$ state and the unusual decay pattern might be explained by
the node structure of the wave functions. Ref. \cite{zhao} also
dealt with the possibility of interpreting $D^+_{sJ}(2632)$ as a
diquark and anti-diquark bound state. Ref. \cite{barnes} suggested
the same scheme for the narrow width of $D^+_{sJ}(2632)$. However
it was pointed out that the relative branching ratio of such a
assignment still disagrees with SELEX experiment \cite{barnes}.

We have inferred from the unusual decay pattern that the narrow
resonance observed by SELEX Collaboration is very probably a four
quark state with the quark content ${1\over 2\sqrt{2}}
(ds\bar{d}+sd\bar{d}+su\bar{u}+us\bar{u}-2ss\bar{s})\bar{c}$. We
have made a systematic analysis of the charm tetraquark states and
have explicitly demonstrated how this unusual decay pattern
occurs.

One of the authors (Y. R. L.) thanks W. Wei for cross-checking the
SU(3) Clebsch-Gordan coefficients. This project was supported by
the National Natural Science Foundation of China under Grant
10375003, BEPC Opening Project, Ministry of Education of China,
FANEDD and SRF for ROCS, SEM.

\end{document}